\documentclass[reviewcopy]{elsarticle}

\usepackage[reviewcopy]{adndt}
\usepackage{longtable}

\usepackage{mathptmx}

\usepackage{amsmath}

\usepackage{amssymb}

\biboptions{square,sort&compress}
\bibpunct[]{[}{]}{,}{n}{}{;}
\begin{document}

\begin{frontmatter}

\journal{Atomic Data and Nuclear Data Tables}


\title{Theoretical energy level spectra and transition data for 4p$^6$4d$^2$, 
4p$^6$4d4f and 4p$^5$4d$^3$ configurations of W$^{36+}$ ion}
  \author{P. Bogdanovich\corref{cor1}}
 \ead{Pavelas.Bogdanovicius@tfai.vu.lt}

  \author{R. Kisielius}

  \cortext[cor1]{Corresponding author.}

   \address{Institute of Theoretical Physics and Astronomy, Vilnius University,
 A. Go{\v s}tauto 12, LT-01108 Vilnius, Lithuania}


\date{09/01/2012} 

\begin{abstract}  
The {\it ab initio} \/ quasirelativistic Hartree-Fock method developed 
specifically for the calculation of spectral parameters of heavy atoms and 
highly charged ions is used to derive transition data for multicharged 
tungsten ion. The configuration interaction method is applied to include 
electron correlation effects. The relativistic effects are taken into account
in the Breit-Pauli approximation for quasirelativistic Hartree-Fock radial 
orbitals. The energy  level spectra, radiative lifetimes $\tau$, Lande 
$g$-factors are calculated for the $\mathrm{4p^64d^2}$, $\mathrm{4p^64d4f}$ and  
$\mathrm{4p^54d^3}$ configurations of W$^{36+}$ ion. The transition wavelengths 
$\lambda$, spontaneous transition probabilities $A$, oscillator strengths $gf$ 
and line strengths $S$ for the electric dipole, electric quadrupole, electric 
octupole  and magnetic dipole transitions among the levels of these 
configurations are tabulated. 

\end{abstract}

\end{frontmatter}

\newpage

\tableofcontents
\listofDtables
\listofDfigures

\section{Introduction}

The tungsten is a very promising constructive material, therefore its spectral 
properties are of increased interest both for experimental and theoretical 
physics \cite{01,02,03,04,05,06}. Although it is difficult to melt and to 
vaporize the metallic tungsten, the tungsten ions with very high degrees of 
ionization are present in various high-temperature devices, including the fusion 
reactors \cite{07,08}. These ions make negative impact on plasma by cooling it. 
Therefore one needs to control continuously the concentration of tungsten ions 
by monitoring the emission spectra of these highly-charged ions. An extensive 
review of the experimental energy level spectra and transition wavelengths for 
tungsten ions of various ionization degrees was presented in \cite{09}. It is 
evident from that review that the data for the tungsten ions having ionization 
degrees of few dozens are insufficient.

It is quite expensive and complicated to study the spectral parameters of 
multicharged ions in experiments. The consistent theoretical studies of tungsten
ions can facilitate the improvement and acceleration of experimental research. 
Current work continues theoretical studies of tungsten ions presented in 
\cite{10,11,12,13, 14,15,16,17}. Similarly to earlier works \cite{11,15,16,17}, 
we investigate further the ions with partially-filled 4d-shell. In present work 
we determine spectral parameters of the W$^{36+}$ ion having two 4d-electrons 
in the ground configuration. The configurations under investigation, 
$\mathrm{4p^64d^2}$, $\mathrm{4p^64d4f}$ and  $\mathrm{4p^54d^3}$, exhibit a 
significantly richer spectral properties comparing with those of previously 
studied configurations $\mathrm{4p^64d}$, $\mathrm{4p^64f}$ and  
$\mathrm{4p^54d^2}$ \cite{16,17} of the W$^{37+}$ ion. We implement a 
quasirelativistic approach (QR) with extensive inclusion of correlation 
effects by exploiting configuration interaction (CI) method as it has been 
performed in our above-mentioned studies.

In Section~\ref{calc} we provide a description of our calculation method. 
Since the adopted approximation completely matches one described in \cite{17} 
for the W$^{37+}$ atomic data investigation, we provide only a brief summary of 
our  method in present work. Obtained results, their accuracy are discussed in 
Section~\ref{result} where they are compared with available experimental data
and theoretical data from other authors.

\section{Calculation method}{\label{calc}}

We use a quasirelativistic approximation in our {\it ab initio} calculations of 
the ion energy level spectra. This approach significantly differs from widely 
used method described in \cite{18}. The main differences arise from our adopted 
set of quasirelativistic Hartree-Fock equations (QRHF) featuring several 
distinctive properties, namely: 

\begin{enumerate}[1.]
\item 
No statistical potentials are used. There are only conventional self-consistent 
field direct and exchange potentials in QRHF.
\item
The finite size of the nucleus is taken into account to determine the potential 
while solving the quasirelativistic equations \cite{19,20}.
\item
The mass-velocity term is divided into two parts - the direct potential and the 
exchange one.
\item
The contact interaction term contains only the nucleus potential derivative in 
the numerator. There are no two-electron potentials in the numerator.
\item
Only the direct part of the potential is included into the denominator of the 
contact interaction with a nucleus term.
\item
The contact interaction with a nucleus is taken into account not only for 
s-electrons, but also for p-electrons with some additional corrections made.
\end{enumerate}

The most complete description of the way to derive the QRHF equations employed 
in this work is given in \cite{21,22}. The methods to calculate the energy 
level spectra in the adopted QR approach were discussed extensively in 
\cite{23}. In current work the main consideration was given to the interaction 
of quasi-degenerate configurations in calculation of electron correlation 
effects. The configurations of this type are arising by a way of one-electron 
and two-electron excitations without the change of the principal quantum number
$n=4$. Therefore all the electrons present in quasi-degenerate configurations 
are described by solutions of QRHF equations. On top of that, the excitation of 
all $n=4$ electrons from the adjusted configurations to virtually-excited 
shells with $n>4$ was considered.

The correlation effects are included within CI approximation with 
virtually-excited electrons described by the transformed radial orbitals (TRO) 
(see \cite{23} for more details). Following methods of \cite{17}, the TRO basis 
for all one-electron orbitals with the main quantum number $5 \leq n \leq 7$ 
was produced for all allowed values of the orbital quantum number $\ell$. By 
establishing such a basis of radial orbitals, one can generate 468 even-parity 
non-relativistic configurations with more than 450 thousand $LS$-terms for the 
adjusting of the ground $\mathrm{4p^64d^2}$ configuration, and 982 odd-parity 
configurations having more than 3384 thousand $LS$-terms to adjust the excited 
configurations $\mathrm{4p^64d4f}$ and $\mathrm{4p^54d^3}$. 

In order to reduce the size of Hamiltonian matrices, we follow the procedure 
described in \cite{24} and remove those admixed configurations within the CI 
expansion of the adjusted configuration wavefunction which have the mean weight
${\bar W}_{\mathrm{PT}} < 2 \cdot 10^{-6}$. The parameter 
${\bar W}_{\mathrm{PT}}$ is determined in the second order of perturbation 
theory (PT):
\begin{equation}
{\bar W}_{\mathrm{PT}}(K_0,K^{\prime}) = 
\frac{\sum_{TLST^{\prime}}(2L+1)(2S+1) 
\langle K_0TLS \Vert H \Vert K^{\prime}T^{\prime}LS \rangle ^2}
{g(K_0) \left( {\bar E}(K^{\prime}) - {\bar E}(K_0) \right)^2}.
\label{eq-w}
\end{equation}
where $\langle K_0TLS \Vert H \Vert K^{\prime}T^{\prime}LS \rangle$ is 
a Hamiltonian matrix element  for interaction between adjusted $K_0$ and admixed 
$K^{\prime}$ configuration $LS$-terms, $g$ is statistical weight of the
configuration $K_0$, ${\bar E}$ are averaged energies of configurations.
More details about selection method parameter ${\bar W}_{\mathrm{PT}}$ were 
presented in \cite{17}.

Using this method, the number of configurations in the CI wavefunction expansion
was reduced to 115 for even-parity states and to 222 for odd-parity states
whereas the accuracy of calculated data had not decreased noticeably. 
By applying methods from \cite{25}, we have included only those $LS$-terms from
admixed configurations which interact with the terms of adjusted configuration.
Consequently, the number of $LS$-terms for even-parity states was reduced to
29725 and for odd-parity states to 621183. Such a reduction allows us to employ
our codes and computers to perform calculations of atomic data for the 
$\mathrm{4p^64d^2}$, $\mathrm{4p^64d4f}$ and $\mathrm{4p^54d^3}$ configurations 
of the W$^{36+}$ ion.

All the calculations were performed in $LS$-coupling. The energy operator was 
determined in quasirelativistic Breit-Pauli approximation \cite{23}. Obtained
multiconfigurational eigenvalues and eigenfunctions were employed to describe 
the energy level spectra and to determine data for the electric dipole (E1), 
electric quadrupole (E2), electric octupole (E3) and the magnetic dipole (M1) 
transitions. We investigated the transitions between the levels of different 
configurations and between the levels of same configurations and hence were 
able to determine the accurate radiative lifetimes $\tau$ of all excited levels.

We used our original computer codes for the data production along with the codes
\cite{26,27,28} adopted for our calculation needs.

\section{Results and discussion}{\label{result}}

The experimental energy level spectra of the W$^{36+}$ ground configuration 
$\mathrm{4p^64d^2}$ given both in the review \cite{09} and in NIST database 
\cite{29} contain a level $^1S_0$ for which the energy was determined from the
semirelativistic calculations but not from measurements. The value of the energy
level $^1G_4$ ($E= 182760$cm$^{-1}$) is marked as a questionable one. We compare 
the experimental energy level spectra of the ground $\mathrm{4p^64d^2}$ 
configuration with results from our calculations and  data from the recent 
multiconfiguration Dirac-Fock (MCDF) calculations \cite{30a} in 
Table~\ref{tableA}. The energy level spectra in \cite{15} was determined in 
relativistic approximation too, and these data are very close to results from 
\cite{30a}. It is interesting to observe that all theoretical calculations 
present the ordering of levels $^1G_4$ and $^3P_1$ which is different from 
the order given in  database \cite{29}. The most probable reason for this 
could be the above-mentioned inaccuracy of level 
$^1G_4$ energy value. It is clear from this Table that theoretical QR energy 
level values agree with the experimental data within approximately $1\%$. 
The comparison of QR and MCDF calculation results indicate that 
quasirelativistic results are closer to experimental data in most cases. This
can be explained by the fact, that only 8 configurations were used in MCDF 
calculations for the correlation effect inclusion, whereas this number was 115
in present work.

\renewcommand{\baselinestretch}{1.0}
\begin{table}[hb!]
\caption{\label{tableA}
Comparison of calculated QR energy levels (in 100 cm$^{-1}$) of W$^{36+}$ with 
available experimental data (Exp) from \cite{29} and theoretical MCDF results
\cite{30a}. 
}
\begin{center}
\begin{tabular}{cccrrrr}
\hline
&
\multicolumn{1}{c}{Relativistic}&
\multicolumn{1}{c}{} & 
\multicolumn{1}{c}{} & 
\multicolumn{1}{c}{Exp}& 
\multicolumn{1}{c}{QR}&
\multicolumn{1}{c}{MCDF} \\
$N$ &
\multicolumn{1}{c}{configuration \cite{30a}}&
\multicolumn{1}{c}{$LS$} & 
\multicolumn{1}{c}{$J$} & 
\multicolumn{1}{c}{\cite{29}}& 
\multicolumn{1}{c}{}&
\multicolumn{1}{c}{\cite{30a}} \\
\hline\noalign{\vskip4pt}
1& 4d$_{3/2}^2$	         & $^3$F& 2	&   0	  &   0 &   0	\\
2& 4d$_{3/2}^2$	         & $^3$P& 0	& 678   & 668 & 648	\\
3& 4d$_{3/2}$ 4d$_{5/2}$	& $^3$F& 3	&1413	  &1411 &1407	\\
4& 4d$_{3/2}$ 4d$_{5/2}$	& $^3$P& 2	&1741	  &1722 &1712	\\
5& 4d$_{3/2}$ 4d$_{5/2}$	& $^3$P& 1	&1846	  &1819 &1814	\\
6& 4d$_{3/2}$ 4d$_{5/2}$	& $^1$G& 4	&1828?	 &1833 &1830	\\
7& 4d$_{5/2}^2$	         & $^3$F& 4	&3085	  &3098 &3084	\\
8& 4d$_{5/2}^2$	         & $^1$D& 2	&3315	  &3304 &3287	\\
9& 4d$_{5/2}^2$	         & $^1$S& 0	&[4070] &4071 &4046	\\
\hline\\
\end{tabular}
\end{center}
\end{table}

The list of 20 most important admixed configurations sorted in decreasing order
of their mean weights ${\bar W}_{\mathrm{CI}}$ is given in Table~\ref{tableB}. 
These mean weights were determined from the eigenfunctions of the ground 
configuration $\mathrm{4p^64d^2}$ obtained after Hamiltonian energy matrix 
diagonalization. The Table gives the configuration numbers $N_{\mathrm{PT}}$ 
describing their order derived from PT. As it is seen from 
Table~\ref{tableB}, the order of configurations matches almost exactly the 
order derived from the initial PT calculation (without the Hamiltonian matrix
diagonalization). We mark in bold those configurations which are present in 
the list of configurations from \cite{30a}. It follows from QR calculations, 
that two quasi-degenerate configurations, 4s$^1$4p$^5$4d$^3$4f$^1$ and 
4s$^2$4p$^4$4d$^2$4f$^2$ which significantly affect the wavefunctions and the 
energy spectra of the adjusted configuration $\mathrm{4p^64d^2}$ are absent 
from that list. It is not possible to compare the influence  of admixed 
configurations having electrons with $n \geq 5$ because the MCDF calculation 
employes the solutions of standard Dirac-Fock equations whereas TRO correspond 
to the solutions of multiconfigurational equations specifically adopted to 
include correlation effects.

\renewcommand{\baselinestretch}{1.0}
\begin{table}[th!]
\caption{\label{tableB}
Admixed configurations with the largest averaged contributions 
${\bar W}_{\mathrm{CI}}$ in the wavefunction of the ground configuration of 
W$^{36+}$.
}
\begin{center}
\begin{tabular}{rlrl}
\hline\\
\multicolumn{1}{c}{$N$} &
\multicolumn{1}{c}{${\bar W}_{\mathrm{CI}}$} & 
\multicolumn{1}{c}{$N_{\mathrm{PT}}$} & 
\multicolumn{1}{l}{Configuration} \\
\hline\noalign{\vskip4pt}
1	 & 9.813E-01	&  1	& $\mathbf{4s^24p^64d^2}$       \\
2	 & 7.402E-03	&  2	& $\mathbf{4s^24p^44d^4}$       \\
3	 & 6.395E-03	&  3	& $\mathbf{4s^24p^54d^24f^1}$   \\
4	 & 9.256E-04	&  4	& 4s$^1$4p$^5$4d$^3$4f$^1$      \\
5	 & 7.681E-04	&  5	& 4s$^2$4p$^4$4d$^2$4f$^2$      \\
6	 & 5.095E-04	&  7	& $\mathbf{4s^14p^64d^3}$       \\
7	 & 4.988E-04	&  6	& $\mathbf{4s^24p^64f^2}$       \\
8	 & 2.183E-04	&  8	& 4s$^2$4p$^5$4d$^1$4f$^1$5g$^1$\\
9	 & 1.663E-04	& 10	& 4s$^2$4p$^4$4d$^3$5g$^1$      \\
10	& 1.552E-04	&  9	& $\mathbf{4p^64d^4}$           \\
11	& 1.294E-04	& 11	& 4s$^2$4p$^4$4d$^2$5p$^2$      \\
12	& 1.284E-04	& 13	& 4s$^2$4p$^5$4d$^1$5p$^1$5d$^1$\\
13	& 1.239E-04	& 14	& 4s$^1$4p$^6$4d$^2$5g$^1$      \\
14	& 1.175E-04	& 15	& 4s$^1$4p$^5$4d$^2$4f$^1$5g$^1$\\
15	& 1.077E-04	& 12	& 4s$^2$4p$^5$4d$^2$5f$^1$      \\
16	& 8.140E-05	& 16	& 4s$^1$4p$^5$4d$^2$5s$^1$5p$^1$\\
17	& 7.232E-05	& 18	& 4s$^2$4p$^5$4d$^2$5p$^1$      \\
18	& 5.969E-05	& 20	& 4s$^1$4p$^6$4d$^1$4f$^2$      \\
19	& 5.446E-05	& 17	& 4s$^2$4p$^4$4d$^3$6d$^1$      \\
20	& 4.925E-05	& 19	& 4s$^2$4p$^4$4d$^2$5g$^2$      \\
21	& 4.766E-05	& 22	& 4p$^6$4d$^2$4f$^2$            \\
\hline\\
\end{tabular}
\end{center}
\end{table}

The ionization energy value for the ground level of  the W$^{36+}$ ion obtained 
by semiempirical methods in \cite{30} is 1.57\,keV, whereas our theoretical 
value derived from the present calculation and the data obtained for W$^{37+}$ 
in the same approach in \cite{16,17} is 1.53\,keV. The database \cite{29} 
contains experimentally measured data from \cite{31} only for one energy level 
$^3G_3$ of the excited configuration $\mathrm{4p^64d4f}$ and gives its value 
$E= 1847100$ cm$^{-1}$. Our quasirelativistic calculations provide the energy 
level value of $E = 1870400$ cm$^{-1}$, the deviation being only slightly higher
than $1\%$. It is reasonable to expect that our theoretical results for other 
energy levels will have a very similar accuracy. 

The energy level spectra of the configurations $\mathrm{4p^64d^2}$, 
$\mathrm{4p^64d4f}$ and $\mathrm{4p^54d^3}$ are presented in 
Table~\ref{table1}. The energy level numbers (indices) $N$ given in the first 
column of Table~\ref{table1} and their total angular momenta $J$ are used 
further in Table~\ref{table2} to describe the transition lines. The Lande 
$g$-factors and the radiative lifetimes $\tau$ for the energy levels of the 
configurations $\mathrm{4p^64d^2}$, $\mathrm{4p^64d4f}$ and 
$\mathrm{4p^54d^3}$ are given in columns 4 and 5.

In Table~\ref{table1} the levels are decribed by their $LS$-coupling function 
squared expansion coefficients $W$ in the multiconfiguration CI expansion 
wavefunction. We give the squared expansion coefficients in percents in 
decreasing order. Only the coefficients $W$ exceeding $5\%$ are presented, 
while the expansion itself is limited to maximum five terms. As one could 
expect, the analysis of expansion  coefficients confirms such a strong mixing 
of configurations $\mathrm{4p^64d4f}$ and $\mathrm{4p^54d^3}$ that it is 
inappropriate to investigate them separately. 

Several high-lying energy levels of the odd-parity configurations 
$\mathrm{4p^64d4f}$ and $\mathrm{4p^54d^3}$ are intermixed with low-lying
energy levels of the first excited even-parity configuration. 
Therefore we include ten lowest levels lying below 2\,700\,000 cm$^{-1}$ in
Table~\ref{table1}. We do not assign level numbers for these energy levels.
Also, we do not provide radiative lifetimes for them since we have not 
calculated transition probabilities for these levels. We want to draw attention
that the admixed-configuration basis adopted for the calculation of the excited
even-parity configuration levels had to be reduced substantially comparing
to the configuration basis used for the configurations $\mathrm{4p^64d^2}$, 
$\mathrm{4p^64d4f}$ and $\mathrm{4p^54d^3}$. While computing energy spectra of
five even-parity configurations, we need to select admixed configurations for
each of them. Due to restrictions in our computer resources we had to 
reduce a number of admixed configurations by applying selection criteria
$2 \cdot 10^{-5}$ which was 10 times larger comparing to other configurations
investigated in current work. Subsequently, the accuracy of these calculated 
data may be lower.

We compare our quasirelativistic M1 emission transition probabilities $A$ 
between the levels of the ground configuration $\mathrm{4p^64d^2}$ with the 
relativistic calculation results from \cite{15,30a} in Table~\ref{tableC}. 
The first two columns give the indices $N$ for the initial (upper) and the 
final (lower) levels given in Table~\ref{tableA}. Likewise in the energy 
spectra case, the agreement of transition probabilities from two 
fully-relativistic calculations is very good. The QR results slightly differ 
from the MCDF data, but the deviations are small and are caused by to some 
degree different use of radial orbitals and an altered basis of admixed 
configurations. Our calculation presents more lines for transitions between 
the configuration $\mathrm{4p^64d^2}$ levels. All their parameters are given in 
Table~\ref{table2}. The performed comparison of our QR calculation results with
the data from the fully-relativistic calculations leads to conclusion that
adopted quasirelativistic approach allows to account for both relativistic and
correlation corrections with sufficient accuracy in the case of investigated
W$^{36+}$ ion.

\renewcommand{\baselinestretch}{1.0}
\begin{table}[ht!]
\caption{\label{tableC}
Comparison of M1 transition probabilities (in s$^{-1}$) calculated in this 
work (QR) with MCDF results from \cite{15,30a} for W$^{36+}$.
}
\begin{center}
\begin{tabular}{lllll}
\hline\\
\multicolumn{1}{c}{$N_i$} &
\multicolumn{1}{c}{$N_f$} &
\multicolumn{1}{c}{QR} & 
\multicolumn{1}{c}{MCDF \cite{15}} & 
\multicolumn{1}{c}{MCDF \cite{30a}} \\ 
\hline\noalign{\vskip4pt}
3	&1	&5.46E+04	&5.35E+04	&5.39E+04\\
4	&1	&2.84E+04	&2.76E+04	&2.77E+04\\
5	&2	&2.04E+04	&2.10E+04	&2.11E+04\\
 	&1	&3.84E+03	&3.76E+03	&3.77E+03\\
7	&3	&6.14E+04	&5.92E+04	&5.95E+04\\
 	&6	&1.59E+04	&1.55E+04	&1.55E+06\\
8	&4	&4.90E+04	&4.77E+04	&4.80E+04\\
 	&5	&2.10E+04	&2.03E+04	&2.04E+04\\
 	&3	&1.30E+04	&1.27E+04	&1.27E+04\\
9	&5	&1.57E+05	&1.51E+05	&1.52E+05\\
\hline\\
\end{tabular}
\end{center}
\end{table}

Calculated line data are presented in Table~\ref{table2}. Here we provide the 
level numbers $N$ and the total angular momenta $J$ for the initial (upper) and 
the final (lower) levels, the type of transition, the transition wavelength 
$\lambda$, the emission transition probability $A$,  weighted oscillator 
strength $gf$ value and the line strength $S$. We present all occuring 
transitions from the excited level to the lower levels in decreasing order of 
transition probabilities but we remove the lines having transition 
probabilitiess 50 times smaller than the strongest line. 

Some transition wavelengths $\lambda$ in \cite{31,31a} were calculated in a 
fully-relativistic approximation with correlation effects included. One can 
see that their wavelengths for the transitions within the ground 
$\mathrm{4p^64d^2}$ configuration agree better with the measured data than
our results do.  But for the most transitions the differences between two
theoretical results are significantly smaller than their deviations from
experimental data. Only one transition from the excited $\mathrm{4p^64d4f}$
configuration level $^3G_3$ to the ground level was identified in experiment.
For this particular transition, our wavelength value is slightly more accurate
than the calculated value from \cite{31a}.

The theoretical wavelength $\lambda$ values and transition probabilities $A$ for heavy
ions in a visible light range were reported in \cite{32}. Only data for one 
transition from the $^1G_4$ to $^3F_3$ level within the ground configuration of 
W$^{36+}$ from the $^1G_4$ to $^3F_3$ level are presented there. Their 
calculated transition wavelength $\lambda = 2244$\,\AA, our calculation gives
the values $\lambda = 2368$\,\AA, and the measured wavelength is 
$\lambda = 2412$\,\AA. Our calculated transition probability $A= 561$ s$^{-1}$ is in a
reasonable (for a week M1 line) agreement with $A= 685$ s$^{-1}$ \cite{32} for 
the transition from the $N=6$ ($J=4$) level to $N=3$ ($J=3$) level.

One can notice from Table~\ref{table2} that electric octupole E3 transition
lines appear between other important lines for the transitions from the level
$N=24$ with $J=6$ of the excited configuration. The electric dipole E1 
transitions from this level to the levels of the ground configuration 
$\mathrm{4p^64d^2}$ are forbidden by selection rules for the total angular
momentum $J$, but the values of E3 transition probabilities are of the same 
order of magnitude as those of the M1 transitions within the same configuration. 
Consequently the lifetime $\tau$ for this level presented in Table~\ref{table1} 
has decreased 2.2 times when the E3 transitions are considered. This particular
level has a maximum possible value of the total angular momentum and it is the
lowest level with $J=6$ in its configuration group. Such a value of $J$ makes
the E1 transitions to the ground level forbidden and leaves only few possible
E2 and M1 transitions inside configuration. Even the reduced value the radiative
lifetime $\tau$ of this levels remains very large. It exceeds by 3 to 4 orders
of magnitude the radiative lifetimes both for the even parity and the odd parity
metastable levels.

There are no other levels in the investigated configurations  of  the W$^{36+}$ 
ion for which the electric octupole transitions make a considerable influence 
also. The preliminary calculations for other tungsten ions with an open 
4d-shell indicate that similar effect is possible. Nevertheless, the levels 
the W$^{37+}$ ion do not display such a pecularities, so the level radiative 
lifetime $\tau$ results tabulated in \cite{17} do not require any correction.

\ack
This work, partially supported by the European Communities under the contract 
of Association between EURATOM/LEI FU07-CT-2007-00063, was carried out within
the framework of the European Fusion Development Agreement.

\clearpage
\clearpage				

\TableExplanation

\bigskip
\renewcommand{\arraystretch}{1.0}

\section*{Table 1.\label{tab1expl} 
Energy levels $E$ (in $10^2$ cm$^{-1}$), Lande $g$-factors, radiative lifetimes
$\tau$ (in $10^{-9}$s) and percentage contributions for  $\mathrm{4p^64d^2}$, 
$\mathrm{4p^64d4f}$ and  $\mathrm{4p^54d^3}$ configurations of W$^{36+}$. 
}

Throughout the Table we present all the eigenfunction components which have
their percentage contributions higher than $5\%$. A number of contributions is
limited to five.

\begin{center}
\begin{tabular}{@{}p{1in}p{6in}@{}}
$N$          & The energy level number (index)\\
$E$          & The level energy in $10^2$ cm$^{-1}$\\
$J$          & The total angular momentum $J$ \\
$g$          & The Lande $g$-factor. The notation aEb means 
                  $\mathrm{a \times 10^{b}}$\\
$\tau$       & The radiative lifetime in $10^{-9}$ s \\
Contribution & The percentage contribution of CSF in level eigenfunction
               A subscript on the left side of intermediate term for 4d$^3$ 
               shell denotes a seniority number $\upsilon$\\
\end{tabular} 
\end{center}


\renewcommand{\arraystretch}{1.0}

\bigskip

\section*{Table 2.\label{tab2expl} 
Parameters of electron transitions between the levels of  
$\mathrm{4p^64d^2}$, $\mathrm{4p^64d4f}$ and  $\mathrm{4p^54d^3}$ 
configurations of W$^{36+}$. 
}

The transitions from the upper level to the lower ones are presented in the
descending order  of the transition probability values $A$. The number of presented 
transitions is limited to those having the transition probability values less than 
50 times smaller comparing to the strongest transition line from the particular 
energy level.

\begin{center}


\end{document}